**Tensor-valued diffusion MRI in under 3 minutes: An initial survey of microscopic anisotropy and tissue heterogeneity in intracranial tumors**

Markus Nilsson[1], Filip Szczepankiewicz[2], Jan Brabec[3], Marie Taylor[1], Carl-Fredrik Westin[2], Alexandra Golby[2], Danielle van Westen[1], and Pia C Sundgren[1,4]

[1] *Clinical Sciences Lund, Radiology, Lund University, Lund, Sweden,*

[2] *Brigham and Women's Hospital, Harvard Medical School, Boston, MA, United States*

[3] *Clinical Sciences Lund, Medical Radiation Physics, Lund University, Lund, Sweden,*

[4] *Lund University Bioimaging Center (LBIC), Lund University, Lund, Sweden*

Corresponding author: Markus Nilsson (markus.nilsson@med.lu.se), Clinical Sciences Lund, Radiology, Lund University, Sweden. Tel: +46 70 25 23 745.

**Abstract**

Purpose: To evaluate the feasibility of a 3-minute b-tensor encoding protocol for diffusion MRI-based assessment of the microscopic anisotropy and tissue heterogeneity in a wide range of intracranial tumors.

Methods: B-tensor encoding was performed in 42 patients with intracranial tumors (gliomas, meningiomas, ademonas, metastases). Microscopic anisotropy and tissue heterogeneity were evaluated by estimating the anisotropic kurtosis ($MK_A$) and isotropic kurtosis ($MK_I$), respectively. An extensive imaging protocol was compared with a faster 3-minute protocol.

Results: The fast imaging protocol yielded parameters with characteristics in terms of bias and precision similar to the full protocol. Glioblastomas had lower microscopic anisotropy than meningiomas ($MK_A$ = 0.29±0.06 versus 0.45±0.08, p = 0.003). Metastases had higher tissue heterogeneity ($MK_I$ = 0.57±0.07) than both the glioblastomas (0.44±0.06 p < 0.001) and meningiomas (0.46±0.06, p = 0.03).

Conclusion: Evaluation of the microscopic anisotropy and tissue heterogeneity in intracranial tumor patients is feasible in clinically relevant times frames.

**Introduction**

Diffusion MRI has long been recognized as useful for the characterization of tumor microstructure. Quantification of the apparent diffusion coefficient (ADC) yields an imaging biomarker linked with tumor cellularity (1-3) and monitoring the response of ADC to treatment can enable early prediction of therapy response (4,5). However, the ADC is sensitive also to a number of mechanisms other than the cellularity (6-8). Therefore several approaches have been proposed to improve the ability of dMRI to characterize tumor microstructure. Some rely on microstructure modelling, where assumptions on the microgeometry of tumors are translated into mathematical models that enable estimation of parameters with an assigned interpretation (e.g. the "intracellular volume fraction"). In practise, such approaches may lead to misleading results when the model assumptions are not met (9,10). Another class of approaches ("representations") provide parameters without predetermined interpretations, but that can be associated with microscopic features of the tissue in a pathology-by-pathology basis. An example of the latter is the diffusional kurtosis imaging (DKI) method (11), which demands imaging protocols with higher b-values than what is required for ADC quantification alone, and provides parameters such as the mean kurtosis (MK). DKI has shown promise in enabling a higher sensitivity to tumor microstructure and an improved ability to predict glioma grade (12-14). However, the biological interpretation of the mean kurtosis is ambiguous in tumours, because it is sensitive to both microscopic diffusion anisotropy and intra-voxel variation in isotropic diffusivity also known as tissue heterogeneity (15).

Tensor-valued diffusion encoding is a novel concept for dMRI that can be used to separate microscopic anisotropy from heterogeneity. Where conventional dMRI encode for diffusion by a single pair of pulsed gradients (16), tensor-valued encoding use gradients that encode for diffusion in more than one direction prior to the image readout (17,18). This enables control of the shape of the so-called b-tensor (19-21). Separation of microscopic anisotropy and heterogeneity is enabled by combining dMRI data acquired with more than one shape of the b-tensor and is thus not possible with just conventional dMRI because it can only generate linear b-tensors (20,22-24). Tensor encoding protocols that support separation of microscopic anisotropy and heterogeneity can be based on so-called double diffusion encoding that permits linear and planar tensor encoding (18,25,26) or continuous gradient waveforms that enables encoding with arbitrary b-tensor shapes, for example, combinations of linear and spherical tensor encoding (23,24,27), or any combination of linear, prolate, spherical, and oblate encoding tensors (19-21). Here, we applied linear and spherical encoding because these can be performed with efficient gradient waveforms that contributes to enhanced SNR (28,29).

Data acquired with multiple b-tensor shapes at high b-values can be analyzed using so-called representations comprising higher-order tensors (20), microstructure models (9), or by using a transform similar to the inverse Laplace transform (30). Here we chose the representation approach, because it yields robust parameters and does not require explicit assumptions on the tumor microstructure. The analysis involves estimation of a fourth-order tensor, similar to the kurtosis tensor in diffusional kurtosis imaging (DKI) (11). The difference is that where the data acquisition in DKI necessitates an assumption of full symmetry of this fourth-order tensor, the inclusion of tensor-valued encoding can be used to relax this assumption to one of major and minor symmetry. In practice, this allows the separation of two invariant components of this fourth-order tensor (20), which holds the information on the microscopic anisotropy and heterogeneity of isotropic diffusivities, respectively (24).

Previous studies including tensor-valued acquisitions were relatively lengthy which may hamper their clinical utility. In this study, we utilized the insights from 'fast DKI' that the relevant components of the fourth order tensor can be estimated with a parsimonious and rapid signal sampling scheme (31,32). The purpose of this study was to demonstrate that tensor-valued diffusion encoding can be performed in just 3 minutes of scan time, and to provide an initial survey of the microscopic anisotropy and tissue heterogeneity in various intracranial tumors.

**Theory**

*Signal expression*

The magnetic-resonance signal ($S$) becomes diffusion-weighted by inducing a dispersion in the phase distribution ($\phi$). In the absence of net flow, the signal can be approximated by only considering the second and fourth cumulants of the phase distribution ($c_2$ and $c_4$, respectively):

$$S/S_0 = \langle \exp(-i\phi) \rangle \approx \exp\left(-\tfrac{1}{2}c_2 + \tfrac{1}{24}c_4\right) \qquad \text{Eq. 1}$$

where $\phi = \gamma \int \mathbf{g}(t) \cdot \mathbf{r}(t)\, dt$, $\gamma$ is the gyromagnetic ratio, $\mathbf{g}(t)$ is the magnetic field gradient, and $\mathbf{r}(t)$ is the position of the spin-bearing particle (33). Assuming the voxel can be subdivided into multiple local and non-exchanging microenvironments in which the diffusion is approximately Gaussian, so that effects of time-dependent diffusion on the time-scales of $\mathbf{g}(t)$ can be neglected, we express the second cumulant as

$$c_2 = 2\, \mathbf{B} : \langle \mathbf{D} \rangle$$

where $\mathbf{B}$ is the b-tensor, $\langle \mathbf{D} \rangle$ is the voxel-average of local diffusion tensors, and ':' denotes the double inner product between two tensors, so that $\mathbf{A} : \mathbf{B} = \sum_{i=1..3, j=1..3} a_{ij} b_{ij}$. For completeness, we note that

$$\mathbf{B} = \int_0^\tau \mathbf{q}(t)^{\otimes 2}\, dt$$

where $\otimes$ denotes the outer product so that $\mathbf{x}^{\otimes 2} = \mathbf{x} \otimes \mathbf{x}$, and

$$\mathbf{q}(t) = \gamma \int_0^t \mathbf{g}(t')\, dt'$$

with $\tau$ being the echo time, and that the conventional b-value is given by the trace of the b-tensor:

$$b = \operatorname{Tr}(\mathbf{B}).$$

The fourth cumulant is given by

$$c_4 = \langle c_4' \rangle + 3(\langle c_2'^2 \rangle - \langle c_2' \rangle^2)$$

where $c_2'$ and $c_4'$ denotes the cumulants of the phase dispersions for the local microenvironments. Since we assume that the phase dispersion in each microenvironment is approximately Gaussian, this means that $c_4'$ is approximately zero. Following this assumption, we see that

$$c_4 = 12\, \mathbf{B}^{\otimes 2} : \mathbb{C},$$

where

$$\mathbb{C} = \langle \mathbf{D}^{\otimes 2} \rangle - \langle \mathbf{D} \rangle^{\otimes 2}$$

is the covariance between the diffusion tensors of the local microenvironments (20). Under these assumptions, the MR signal is given by

$$S/S_0 \approx \exp\left(-\mathbf{B} : \langle \mathbf{D} \rangle + \tfrac{1}{2} \mathbf{B}^{\otimes 2} : \mathbb{C}\right) \qquad \text{Eq. 2}$$

The voxel-average diffusion tensor $\langle \mathbf{D} \rangle$ has 6 independent elements and the fourth-order tensor $\mathbb{C}$ has 21 independent parameters because it has major and minor symmetry (20). Methods that supports calculations with fourth-order tensors can be found in the multidimensional diffusion MRI toolbox[1] (34).

*Powder averaging*

In order to simplify estimation of the relevant properties of the tensor covariance, we can utilize so-called powder averaging where the signal is averaged across a number of rotations of the b-tensor. In this section, we assume the b-tensors to be cylinder symmetric. After averaging, the second and fourth cumulants are given by

$$c_2 \approx 2\, \langle \mathbf{B} \rangle : \langle \mathbf{D} \rangle_\mathrm{I}$$

---

[1] https://github.com/markus-nilsson/md-dmri/tree/master/tools/tensor_maths

and

$$c_4 \approx 12 \langle \mathbf{B}^{\otimes 2} \rangle : \mathbb{C}_I$$

where $\langle \mathbf{B} \rangle$ and $\langle \mathbf{B}^{\otimes 2} \rangle$ are averages of $\mathbf{B}$ and $\mathbf{B}^{\otimes 2}$ across the b-tensor rotations, and $\langle \mathbf{D} \rangle_I$ and $\mathbb{C}_I$ are isotropic second and fourth-order tensors. The goal is to include a sufficient number of rotations so that

$$\langle \mathbf{B} \rangle \approx b \tfrac{1}{3} \mathbf{I} \qquad \text{Eq. 3}$$

where $\mathbf{I}$ is the second order identity tensor ($\delta_{ij}$), and

$$\langle \mathbf{B}^{\otimes 2} \rangle \approx b^2 \left( \mathbb{I}_I + b_\Delta^2 \tfrac{2}{5} \mathbb{I}_A \right) \qquad \text{Eq. 4}$$

where and $\mathbb{I}_I$ and $\mathbb{I}_A$ are two isotropic but orthogonal fourth-order tensors ($\mathbb{I}_I : \mathbb{I}_A = 0$). The parameter $b_\Delta$ describes the shape of the b-tensor, is unitless and has a value of –1/2 for planar b-tensor encoding, 0 for spherical b-tensor encoding, and 1 for linear tensor encoding. The isotropic tensors are given by $\mathbb{I}_I = \left(\tfrac{1}{3}\mathbf{I}\right)^{\otimes 2} = \tfrac{1}{9}\delta_{ij}\delta_{kl}$ and $\mathbb{I}_A = \mathbb{I}_T - \mathbb{I}_I$ where $\mathbb{I}_T = \tfrac{1}{6}\left(\delta_{ik}\delta_{jl} + \delta_{il}\delta_{jk}\right)$. In Westin et al (2016), $\mathbb{I}_I$, $\mathbb{I}_A$, and $\mathbb{I}_T$ were denoted $\mathbb{E}_{\text{BULK}}$, $\mathbb{E}_{\text{SHEAR}}$, and $\mathbb{E}_{\text{ISO}}$, respectively.

Under these conditions, the powder averaged signal will be given by

$$S/S_0 \approx \exp\left(-b\langle d_I \rangle + \tfrac{1}{2}b^2[V_I + b_\Delta^2 V_A]\right) \qquad \text{Eq. 5}$$

because

$$\langle d_I \rangle = \langle \mathbf{D} \rangle_I : \tfrac{1}{3}\mathbf{I}$$
$$V_I = \mathbb{C}_I : \mathbb{I}_I = \langle d_I^2 \rangle - \langle d_I \rangle^2$$
$$V_A = \mathbb{C}_I : \tfrac{2}{5}\mathbb{I}_A$$

where $\langle d_I \rangle$ is the voxel-averaged isotropic diffusivity (also referred to mean diffusivity in the context of DTI and DKI), and $V_I$ and $V_A$ are the intra-voxel variances in apparent diffusivities due to isotropic heterogeneity and microscopic anisotropy, respectively (20,24). Note that the average eigenvalue variance of local diffusion tensors, often denoted $\langle V_\lambda \rangle$, is related to $V_A$ according to $\frac{2}{5}\langle V_\lambda \rangle = V_A$.

*A minimal protocol*

Our goal was to enable rapid estimation of the four unknowns in Eq. 5: $S_0$, $d_I$, $V_I$, and $V_A$. At the very minimum, this requires four measurements with different b-tensors ("hypershells") with at least three distinct b-values (e.g. $b = 0$, 1 and 2 ms/μm$^2$) and two shapes of the b-tensor (e.g. $b_\Delta^2 = 0$ and 1). In the absence of voxel-scale anisotropy, four measurements with for example $[b, b_\Delta] = [0, 0]$, $[1, 0]$, $[2, 0]$, and $[2, 1]$, with $b$ in units of ms/μm$^2$, could thus enable estimation of the four parameters in Eq. 5.

In the presence of voxel-scale anisotropy, each hypershell would have to be performed with a sufficient number of rotations of the b-tensors ("directions") to provide an accurate powder average. From theory, we know this number to increase with the b-value (35). For low b-values, rotation invariance is obtained by fulfilling Eq 3. This can be done with a single spherical b-tensor or by averaging over three linear b-tensors:

$$\langle \mathbf{B} \rangle = \frac{1}{3}\begin{bmatrix} b & 0 & 0 \\ 0 & 0 & 0 \\ 0 & 0 & 0 \end{bmatrix} + \frac{1}{3}\begin{bmatrix} 0 & 0 & 0 \\ 0 & b & 0 \\ 0 & 0 & 0 \end{bmatrix} + \frac{1}{3}\begin{bmatrix} 0 & 0 & 0 \\ 0 & 0 & 0 \\ 0 & 0 & b \end{bmatrix} = b\frac{1}{3}\mathbf{I}$$

At higher b-values, we need to fulfil Eq. 4 to obtain rotation invariance. This can be obtained by averaging over the six directions defined by the icosahedral sampling scheme, because

$$\mathbb{I}_I + \frac{2}{5}\mathbb{I}_A = \frac{1}{6}\sum_{i=1}^{6} \boldsymbol{n}_i^{\otimes 4}$$

where $\boldsymbol{n}_1 = [0 \quad c_1 \quad c_2]$, $\boldsymbol{n}_2 = [0 \quad c_1 \quad -c_2]$, $\boldsymbol{n}_3 = [c_1 \quad c_2 \quad 0]$, $\boldsymbol{n}_4 = [c_1 \quad -c_2 \quad 0]$, $\boldsymbol{n}_5 = [c_2 \quad 0 \quad c_1]$, and $\boldsymbol{n}_6 = [-c_2 \quad 0 \quad c_1]$, where $c_1 = \sqrt{(5 - 5^{1/2})/10}$, and $c_2 = \sqrt{(5 + 5^{1/2})/10}$ (36). Previous studies have shown that a similar rotation invariance is possible also by averaging across nine custom directions (37). In practise, just six directions may not be sufficient due to the influence of higher-order terms (e.g. b$^3$ terms) (38). For the purpose of rotation invariance, however, we have previously shown six directions to be sufficient for an accurate powder average up to moderate b-values ($b \cdot d_I < 2$) if the system of interest has a low voxel-level anisotropy (FA < 0.5) (29). Fortunately, this is the case for most tumours.

In summary, the theoretical analysis shows that rotation-invariant estimates of the four model parameters of interest can be obtained in a voxel with low to moderate voxel-level anisotropy with just nine measurements: three with spherical b-tensors having $b$ = 0, 1, and 2 ms/µm² and six with linear b-tensors and $b$ = 2 ms/µm² played out along the icosahedral sampling scheme. In practise, more measurements may be prefered to improve the signal to noise ratio (SNR).

*Microstructure measures*

From the parameters in Eq. 5, we define the two microstructure measures that we will focus on in this study, which we refer to as the isotropic and anisotropic kurtosis ($MK_I$ and $MK_A$, respectively), defined by

$$MK_I = 3\,V_I/\langle d_I \rangle^2$$

and

$$MK_A = 3\,V_A/\langle d_I \rangle^2 \ .$$

The sum of these parameters yields the total mean kurtosis ($MK_T = MK_I + MK_A$), which is similar but not identical to the kurtosis obtained in diffusional kurtosis imaging. The dissimilariy stems from the powder averaging operation applied in the present analysis.

We have previously demonstrated that $MK_A$ and $MK_I$ capture microstructure features of tumors via quantitative analysis of histological images (15). Results showed that the $MK_I$ was associated with heterogeneity within the voxel of the cell density, whereas the $MK_A$ was associated with the average cell shape within the voxel.

**Methods**

*Acquisition protocol*

Imaging was performed on a 3T MAGNETOM Prisma with a 20-channel head coil array (Siemens Healthcare, Erlangen, Germany). Morphological imaging was performed with a T1-weighted 3D-MPRAGE (Magnetization Prepared Rapid Acquisition Gradient Echo pre- and post-intravenous Gadolinium administration and a FLAIR (Fluid-attenuated inversion recovery) sequence. Diffusion-weighted images were acquired prior to the administration of Gadolinium with a prototype spin-echo sequence that enables diffusion encoding with arbitrarily shaped b-tensors. Imaging was performed with TE = 80 ms, TR = 3.2 s, FOV = 230×230 mm$^2$, slices = 21, resolution = 2.3×2.3×2.3 mm$^3$, iPAT = 2 (SENSE), and partial-Fourier = 6/8. Tensor encoding was performed using asymmetric gradient waveforms that were optimized to minimize TE using a constrained optmisation approach described in (28) and available at https://github.com/jsjol/NOW. The optimization used the following settings: "max norm", heat dissipation factor 0.5, and a slew rate limit of 50 T/m/s, to comply with duty cycle and peripheral nerve stimulation limits. A short TR was enabled by limiting the number of slices. The resulting 5-cm-thick imaging volume was positioned across the lesion of interest by the radiographers based on images from previous examinations (CT or MRI).

*Evaluation of the tensor encoding protocol*

The dMRI protocol comprised four b-values ($b$ = 0.1, 0.7, 1.4 and 2.0 ms/μm$^2$) acquired in 3, 3, 6, and 6 directions for the linear tensor encoding, and with 6, 6, 10, 16 averages of the spherical tensor encoding, respectively. This resulted in an acquisition time of 3 minutes. Previous b-tensor encoding protocols has featured more than six directions at the maximum b-value in order to ensure that a rotation-invariant powder-averaged signal could be obtained (15,29). The accuracy of powder averaging with a limited number of directions was analyzed by acquiring extra data in a volunteer using three different diffusion protocols referred to as the 'full', 'subsampled', and 'optimized' protocols. All protocols comprised the same four b-values ($b$ = 0.1, 0.7, 1.4 and 2.0 ms/μm$^2$), but were applied in 6, 6, 10, and 16 directions for the full protocol and in 3, 3, 6, and 6 directions for the subsampled and optimized protocols. All protocols sampled STE signals with 6, 6, 10, and 16 averages for the different b-values. The directions in the full protocol were obtained by the so-called electrostatic repulsion algorithm (39). The directions in the subsampled protocol were selected from the full protocol to be as spread out across the sphere as possible. The optimized protocol was the one used in the full study, and for that protocol the three directions used for the lower b-values were orthogonal, and the six directions used at higher b-values were selected from an icosahedral sampling scheme (that intrinsically minimize the electrostatic repulsion). The accuracy of MD, $MK_A$, and $MK_I$ was then assessed by investigating the difference between the two shorter protocols (subsampled and optimized) and the full protocol.

The accuracy and precision were also investigated by simulations of a system comprised of cylinder-symmetric diffusion tensors with an axial diffusivity of 2.0 μm²/ms and a radial diffusivity of 0.2 μm²/ms. The diffusion tensors were aligned along a given direction with a small orientation dispersion corresponding to an angular standard deviation of 15 degrees. Noise was added to the simulated signal so that it followed a Rice distribution, with a noise-free magnitude given by the true signal, and a noise level corresponding to SNR = 40 at $b = 0$. The noisy signal was then powder-averaged, and used estimate MD, $MK_A$, and $MK_I$. The process was repeated for 1000 random rotations of the diffusion tensors, and the mean and standard deviation of the three parameters were computed for each of the three protocols. For a protocol with an accurately determined powder averaged signal, the standard deviation ($\sigma$) would represent only noise, whereas it would be higher for a protocol with suboptimal directions due to a variable rotation-dependent bias. Assuming the two sources or error are uncorrelated, we can express this as $\sigma^2 = \sigma^2_{rotation} + \sigma^2_{noise}$. Simulations performed without noise thus allowed separate estimation of $\sigma^2_{rotation}$, and thus the two terms were reported separately.

*Subjects*

Patients were recruited from those scheduled for a clinical MRI due to a suspected or recurrent brain lesion, and were enrolled after giving informed consent. The study was approved by the Ethical Review Board in Lund, Sweden. During the period between March 2017 and August 2018, 35 patients with intracranial tumors were enrolled. Out of those, 22 had gliomas (13 glioblastoma, 4 astrocytoma, 2 oligoastrocytoma, 1 oligodendroglioma, 1 glioma, 1 brain stem glioma), 5 meningiomas, 1 hemangiopericytoma, 6 brain metastases (3 with primary breast tumours, 2 with primary lung tumours, and 1 with rectal cancer), and 1 pituitary adenoma. Of the gliomas, 12 had undergone surgery prior to imaging. Another 6 patients were scanned but were not included in the study because there were no lesions visible in the contrast-enhanced T1W images (5 cases) or the lesion was close to regions with strong susceptibility artefacts such as parts of the frontal lobe close to the the petrous apex (1 case).

*Image post processing*

The diffusion-weighted images were processed in three steps. The first step aimed at correction of motion and image distortions from eddy currents, and included registration of the diffusion-weighted volumes to extrapolated references using ElastiX (40). The use of extrapolation-based references has been shown to be necessary for accurate registration of high b-value data (41). In the second step, all volumes were smoothed by a three-dimensional Gaussian kernel with a standard deviation of 0.4 voxels. In the third step, parameter maps were obtained by fitting the mean diffusivity ($\langle d_I \rangle$) and the isotropic and anisotropic diffusion variances ($V_I$ and $V_A$) to the data using Eq. 5. The fitting was performed by linear least squares fitting of the log signal, while correcting for heteroscedasticity. Once these parameters were estimated, the isotropic and anisotropic kurtosis components were computed. In addition to these steps, the post-Gd T1W image volumes were registered to the diffusion-weighted volumes in order to enable the tumor definition for the quantitative analysis. All post processing was performed using the multidimensional diffusion MRI toolbox (34), which is implemented in Matlab (The MathWorks, Natick, MA, USA) and available at https://github.com/markus-nilsson/md-dmri.

For one subject, perfusion maps of the relative cerebral blood volume (rCBV) were calculated using Nordic ICE (NordicNeuroLab, Bergen, Norway) from data acquired with dynamic susceptibility contrast acquisition with a time resolution of 1.5 seconds, using a single-shot gradient echo EPI-gradient sequence and a spatial resolution of 1.7×1.7×6.0 mm$^3$ and an echo time of 28 ms. The maps were computed using truncated singular value decomposition and were leakage-corrected with Boxerman and gamma fitting, and were coregistered with the diffusion data.

*Quantitative analysis*

Regions of interest (ROIs) were drawn in the contrast-enhancing regions on the post-Gd T1W images, excluding apparently necrotic parts where $\langle d_I \rangle > 2$ μm$^2$/ms. Parts of the images affected by image artefacts due to for example insufficient suppression were also excluded. In addition, ROIs were drawn in normal-appearing frontal white matter to characterize normal-appearing white matter. Data were then obtained from all subjects except those with glioblastoma, extensive edema, or an imaging slab that did not cover frontal white matter. Values of the mean diffusivity ($\langle d_I \rangle$), the microscopic anisotropy (MK$_A$) and the tissue heterogeneity (MK$_I$) were obtained for each ROI, and basic descriptive statistics were calculated.

**Results**

Figure 1 shows a comparison of the three sampling protocols: the first was a full protocol (5 minutes long) while the second and third were shorter and referred to as the subsampled and optimized protocols (both 3 minutes long). Numerical simulations showed estimated MD values close to the expected value of 0.8 $\mu m^2$/ms for all three protocols, with average values (standard deviation) of 0.78 (0.03), 0.78 (0.07), and 0.78 (0.04) $\mu m^2$/ms, respectively. For $MK_A$, the expected value was 1.35 and there was a substantial bias for all three protocols, with average (std) values of 1.10 (0.12), 1.14 (0.28), and 1.09 (0.15). A similar level of bias was found for $MK_I$, with average (std) values of –0.12 (0.18), –0.17 (0.42), and –0.09 (0.22), compared with the expected value of zero. Bias of this magnitude is expected when using truncated cumulant expansions for the data analysis (38,42). The parameter uncertainty, represented by standard deviations, had two origins: noise, and incomplete rotation invariance that contributed to random errors due to the random rotation applied to the synthetic sample in the simulations. For both the full and the optimized protocols, the variation due to rotation was substantially smaller than that due to noise, whereas for the subsampled protocol it was larger (Figure 1). Corresponding results were found in the bias maps from the volunteer measurement, where the subsampled protocol showed location-dependent errors versus the full protocol in all parameters, whereas the errors from the optimized protocol appeared negligible. In summary, the full and the short protocols exhibited similar characterstics, whereas the naively subsampled protocol suffered from high variance due to rotation-induced errors.

Figure 2 shows post-Gd T1W and FLAIR images and maps of the mean diffusivity, microscopic anisotropy and tissue heterogeneity in four different types of brain tumors. Data were obtained with the optimized protocol. Within the contrast enhancing parts of the glioma and the metastasis, we note a low but non-zero microscopic anisotropy indicating the presence of some residual white matter. Parts of the enhancing lesions also displayed an elevated tissue heterogeneity. In the glioma and brain metastasis patients, edema surrounded the contrast enhancing lesions. This region exhibited elevated mean diffusivity, reduced microscopic anisotropy, and a moderately increased tissue heterogeneity. The pituitary adenoma and the meningioma both had higher microscopic anisotropy than what was observed in the glioma and the metastasis, indicating the presence of elongated cell structures within these tumors. The pituitary adenoma differed from the meningioma in terms of its tissue heterogeneity, which was clearly elevated.

In some patients, parts of the tumor edges showed exceptionally high tissue heterogeneity (Figure 3). This parameter depends on data acquired with isotropic diffusion weighting (spherical tensor encoding), and regions with high heterogeneity also had a conspicuous contrast in the raw signal data (Figure 4). The increase in image contrast with high b-value spherical encoding compared to conventional linear tensor encoded data is particularly striking.

Follow-up examinations were available for one patient. Figure 6 shows the temporal evolution of the morphological images and the diffusion parameter maps. All diffusion parameter maps were consistent across time on the side contralateral to the lesion, except for some image artefacts. Changes on the maps of the mean diffusivity and microscopic anisotropy on the side ipsilateral to the lesion were aligned with the changes of the edema. The tissue heterogeneity was elevated at baseline, but was gradually reduced at later time points. Elevated tissue heterogeneity co-occurred with low relative blood volumes.

Finally, average parameter values across the tumors are displayed in Figure 6, categorised by tumor type (glioma, glioblastoma, metastasis and meningimoa) and compared with normal-appearing white matter. The glioma group included a diverse set of tumours, which manifested as a high variability between the tumours in this group. All tumors except a two gliomas had higher mean diffusivity, lower microscopic anisotropy, higher tissue heterogeneity, and lower total kurtosis than the normal-appearing white matter. Glioblastomas had lower average microscopic anisotropy than meningiomas ($MK_A$ = 0.29±0.06 versus 0.45±0.08, $p$ = 0.003, Ranksum test). Metastases had higher tissue heterogeneity ($MK_I$ = 0.57±0.07) than both the glioblastomas (0.44±0.06 $p$ < 0.001, Ranksum test) and meningiomas (0.46±0.06, $p$ = 0.03, Ranksum test).

**Discussion**

In this study, we demonstrated that tensor-valued diffusion encoding can be performed in just three minutes to quantify microscopic anisotropy and tissue heterogeneity in brain tumors. In constrast to normal-appearing white matter which exibited high microscopic anisotropy and low tissue heterogeneity, the tumors exhibited low to intermediate microscopic anisotropy and low to high tissue heterogeneity, with the specific characteristics depending on tumor type. Considerable variation was also found within the tumors. Differences in microscopic anisotropy between the glioma and meningioma groups were in line with previous investigations showing that meningiomas contain more microscopically anisotropic tissue (15). High microscopic anisotropy was also found in the pituitary tumor, suggesting that the tumor comprised elongated spindle cells common for e.g. pituicytomas (43), however we reserve final interpretation until $MK_A$ can be associated to structural anisotropy from histology in a larger sample of this type of tumor. Moreover, substantially elevated high tissue heterogeneity were found in some tumors. The biological interpretation of this is unclear, but indicates a high variation of the diffusivity within the voxel. Although speculative, we hypothesise that this could be caused by partial necrosis within the voxel, meaning that some parts of the voxel has high cell density and thus low diffusion whereas others are necrotic with high diffusivity. The co-occurance of low relative blood volumes and high tissue heterogeneity could support this hypothesis (Figure 5). Further investigations are necessary to determine the association between these novel diffusion parameters and pathology.

Previous imaging protocols with b-tensor encoding were longer than the one used in the present study. The acceleration relied on four factors: optimized gradient waveforms, limited number of directions, limited slice coverage, and sample balancing. Gradient waveforms were optimized to make use of all the encoding time available in a spin echo sequence (28). Thus, the gradient waveforms were asymmetric in contrast to the symmetric approach taken in some previous papers (15,20). The limited number of directions made it possible to reduce the total scan time, and by using a combination of directions that provide a balanced sample of the fourth-order tensor, and was inspired by previous papers on 'fast DKI' (31,32). However, due to the limited number of directions (six) we expect a slight rotation-dependent bias in white matter signal with high orientation coherence, but in tumors this bias should be negligible due to their low voxel-level anisotropy (FA) (29). The results in Figure 1 indicates that the orientation-dependent bias in the short protocol is indeed small. The third factor contributing to a faster protocol was the use of a limited imaging slab. This allowed a shorter repetition time, and thus a shorter total scan time. The shorter scan time also led to a lower total heat load of the gradient coil, which in turn could be used to shorten the repetition time further while still respecting duty cycle limits. Although using a limited number of slices also reduce the field of view, this limitation can be addressed by simultaneous multislice acquisitions (44). Finally, high quality maps were enabled by an adapted distribution of samples. In previous works (29), we observed that $MK_I$ had a lower precision than $MK_A$ when acquiring an equal number LTE and STE volumes. Thus, this protocol featured a higher relative fraction of STE volumes. It should be noted, however, that the protocol design was the result of an experience-based act of balancing a number of factors that influenced scan time and parameter precision. Future work could explore formal means of protocols optimization to improve parameter precision (45,46). Such optimization does not generally address the parameter bias reported in Figure 1. This bias is expected when higher order terms affect the acquired data, but the analysis is truncated to second cumulant (38). Finding an optimal protocol requires balancing bias and precision (42), and here we prioritised precision over bias (with the exception of minimising rotation-dependent bias by using a optimised protocol).

We acknowledge three main limitations of the current study. First, the presence of artefacts due to concomitant fields may have led to parameter bias due to the use of asymmetric waveforms (47). Waveforms can today be optimized to mitigate this effects (47), however, this project was initiated before such waveforms were available and may have resulted in a minor bias towards higher values of the microscopic anisotropy. This should be addressed in futures studies. Second, the linear and spherical tensor encoding were performed using gradient waveforms with different timings, so the difference between the acquisitions was not only in the shape of the b-tensor, but also in the effective diffusion time (38,48,49). In the protocol design process, we tried to minimize this difference by making the diffusion time of the linear tensor encoding as short as possible, but the remaining timing difference could have resulted in parameter bias. In healthy white and gray matter, the time-dependence of the diffusion is negligible between approximately 10 and 250 ms (50,51), but whether this is true in all of the tumors investigated remains to be tested. Third, some patients were investigated before surgery and others after. This may have affected the parameters. To test whether the novel diffusion parameters can contribute with diagonstically relevant information, future studies should do imaging prior to treatment.

Future work can utilize this short protocol to test clinically related questions such as whether separation of the two diffusional kurtosis components ($MK_A$ and $MK_I$) can increase the performance of glioma grade discrimination over the total kurtosis ($MK_T$) alone (52), enable mapping of meningioma consistency (53,54), monitor or treatment prediction, or to correlate imaging and histological analysis of biopsies to elucidate the microstructural underpinnings of the observed contrasts.

**Acknowledgements**

This study was supported by grants from Swedish Research Council (2016-03443, 2016-02199-3), Swedish Cancer Society (CAN 2016/365), and the Crafoord Foundation (grant no. 20160990), and Random Walk Imaging AB (grant no. MN15) The funding sources had no role in the design and conduct of the study; in the collection, analysis, interpretation of the data; or in the preparation, review, or approval of the manuscript.

**Conflicts of interest**

MN declares research support from and ownership interests in Random Walk Imaging (formerly Colloidal Resource), and patent applications in Sweden (1250453-6 and 1250452-8), USA (61/642 594 and 61/642 589), and PCT (SE2013/ 050492 and SE2013/050493). FS has been employed at Random Walk Imaging.

# Figures

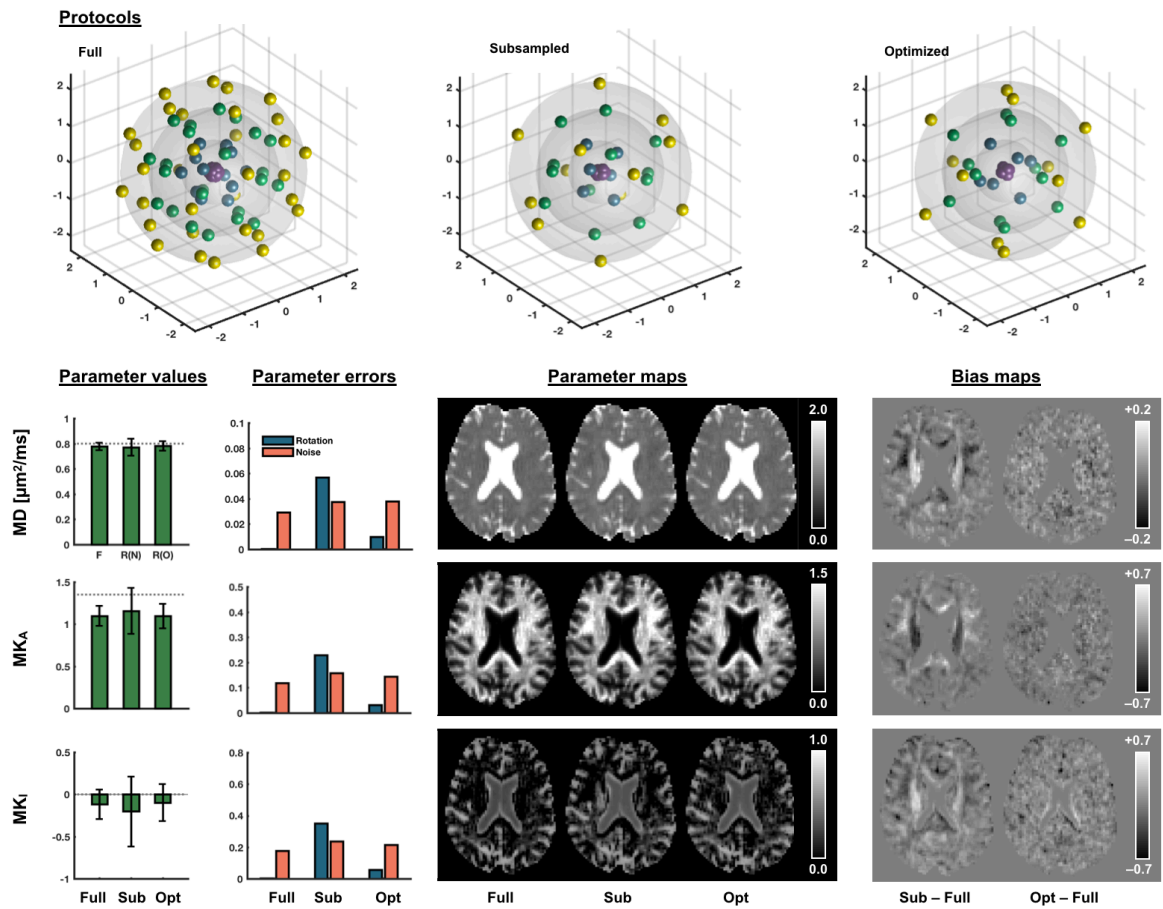

**Figure 1. Accuracy and precision for the different imaging protocols.** The top row visualizes the b-value shells and directions of the linear tensor encoding part of the three protocols. The full protocol featured more directions than the subsampled and optimized protocols. The first columns of the second row show average parameter values (error bars show standard deviations) from the numerical simulation with the different protocols. The dashed line represents the expected values. The second column shows parameter errors from the same simulation. The third and fourth columns show parameter and bias maps, respectively, from a healthy volunteer. The short optimized yielded results similar to the long full protocol, whereas this was not true for the subsampled protocol.

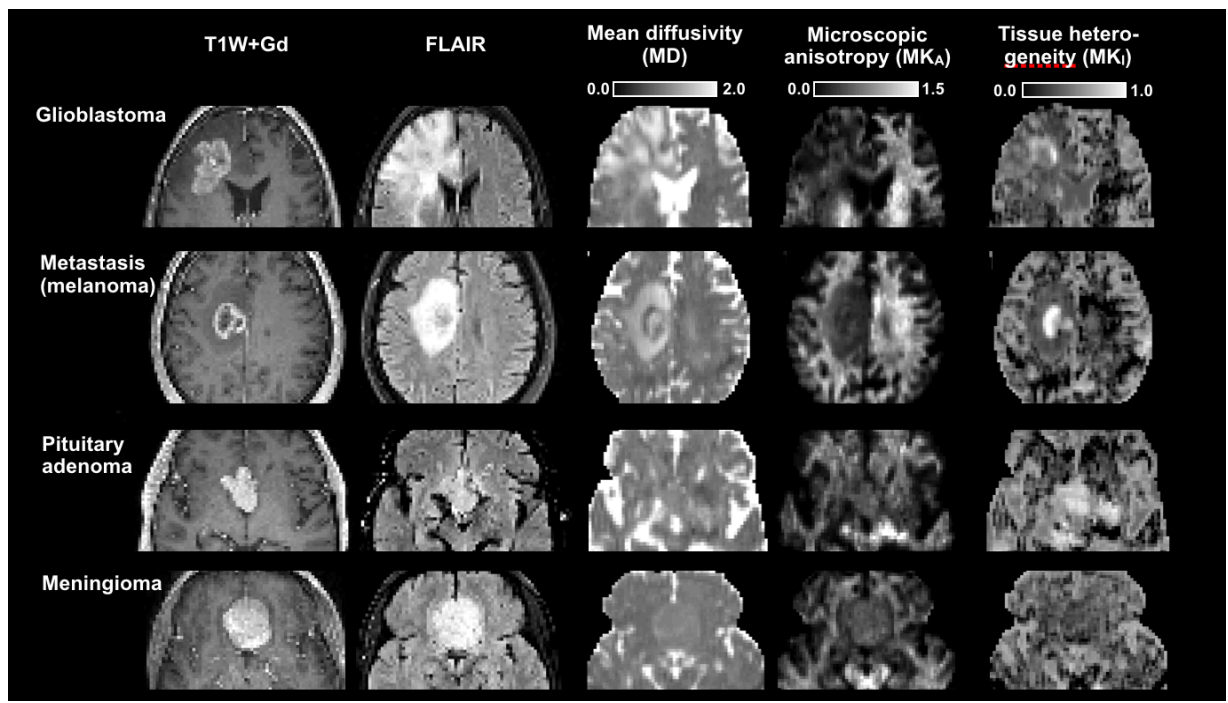

**Figure 2. Morphology and microstructure parameter maps in four brain tumor types.** Both the glioblastoma and brain metastasis exhibited low microscopic anisotropy, whereas the pituitary adenoma and the meningioma exhibited higher microscopic anisotropy. All tumors in this panel, except the meningioma, displayed regions with markedly elevated tissue heterogeneity (isotropic kurtosis).

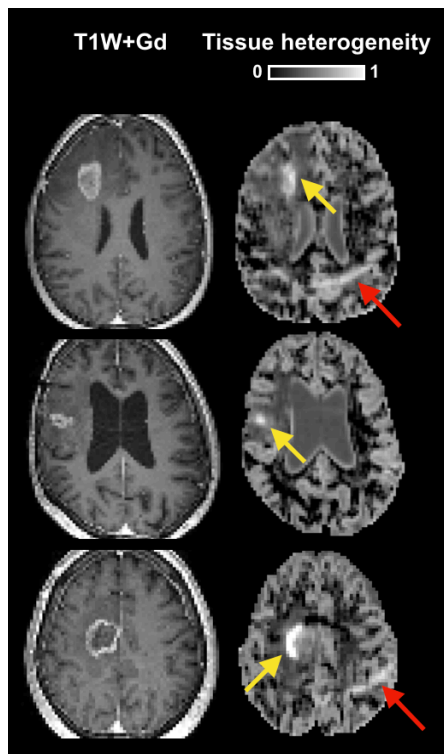

**Figure 3. Examples of tumors with high tissue heterogeneity.** Cases include two glioblastomas (top rows) and one brain metastasis (bottom row). The yellow arrows points to regions with exceptionally high tissue heterogeneity. The red arrow points to an artefact caused by insufficient fat supression.

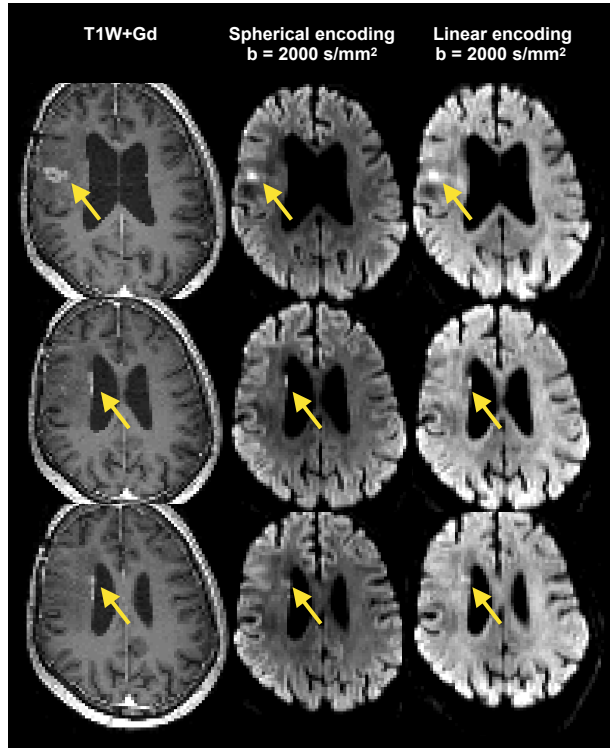

**Figure 4. Illustration of stronger image contrast with spherical compared with linear tensor encoding.** Yellow arrows point to the locations with high gadolinium load. These regions were associated with high contrast in the high b-value image with spherical encoding (middle clumn), whereas a lower contrast was observed when using linear encoding (right column). Identical windowing was applied to the spherical and linear encoding data. Data represent consecutive slices from one patient. The images indicates that tissue anisotropy in surrounding tissue can obfuscate regions of dense tumor tissue, which can reduce detectability.

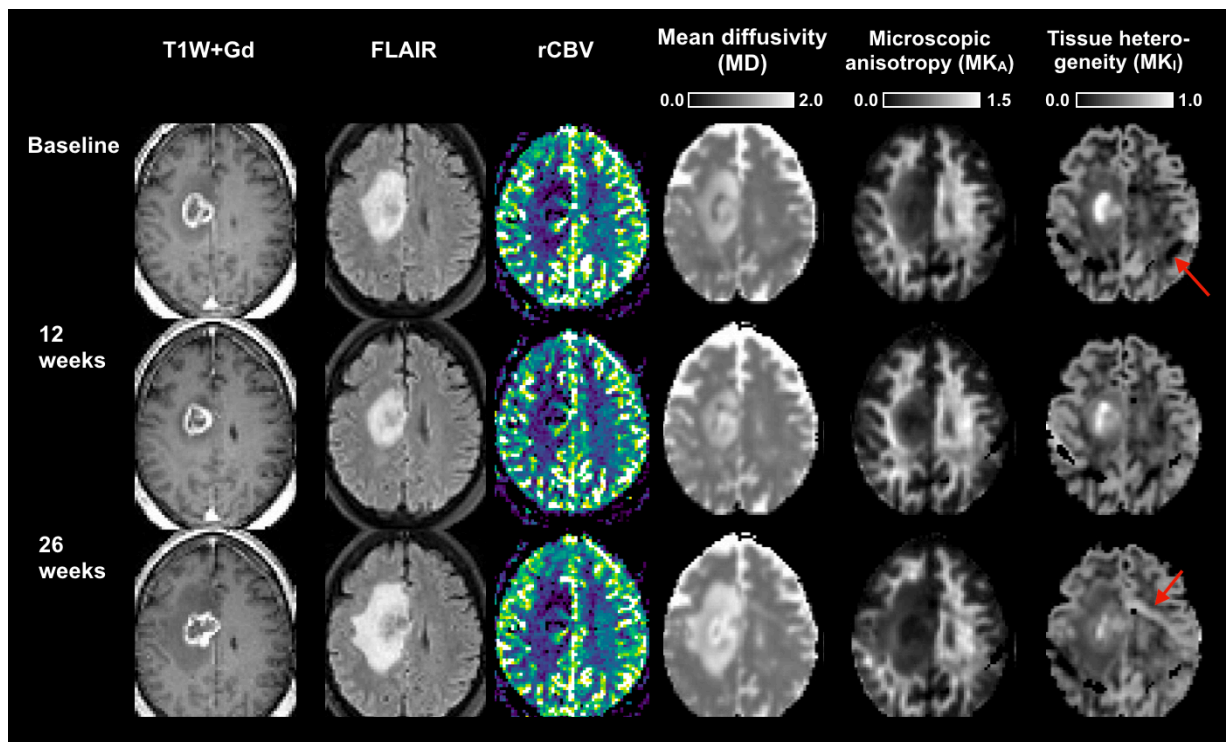

**Figure 5. Temporal evolution of the morphological images, relative cerebral blood volume (rCBV), and the diffusion parameter maps in a brain metastasis patient.** Contrast enhancement is seen on all of the T1W+Gd images. The edema seen on the FLAIR image contracts week 12 but expands week 26. Parts of the tumour shows consistently elevated relative blood volume. All diffusion parameter maps were consistent across time on the side contralateral to the lesion, except for some image artefacts (red arrow). Changes on the maps of the mean diffusivity and microscopic anisotropy on the side ipsilateral to the lesion were aligned with the changes of the edema. The tissue heterogeneity was elevated at baseline, but was gradually reduced at later time points.

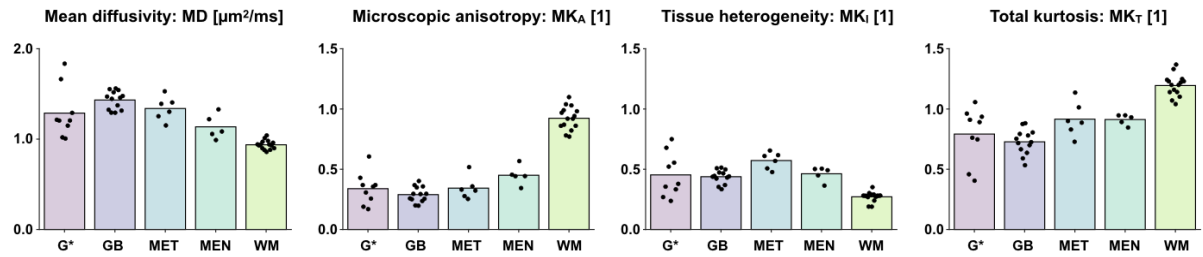

**Figure 6. Overview of parameters in tumors, compared with white matter (WM)**. The tumours were categoriesed by type: glioma excluding glioblastoma (G*), glioblastoma (GB), metastasis (MET), menimgimoa (MEN). Bars show average values, whereas black dots show values from individual patients (averaged across the ROIs).

# References

1. Sugahara T, Korogi Y, Kochi M, et al. Usefulness of diffusion-weighted MRI with echo-planar technique in the evaluation of cellularity in gliomas. J Magn Reson Imaging 1999;9:53–60.

2. Chenevert TL, Stegman LD, Taylor JM, Robertson PL, Greenberg HS, Rehemtulla A, Ross BD. Diffusion magnetic resonance imaging: an early surrogate marker of therapeutic efficacy in brain tumors. J. Natl. Cancer Inst. 2000;92:2029–2036.

3. Chen L, Liu M, Bao J, Xia Y, Zhang J, Zhang L, Huang X, Wang J. The correlation between apparent diffusion coefficient and tumor cellularity in patients: a meta-analysis. Hess CP, editor. PLoS One [Internet] 2013;8:e79008. doi: 10.1371/journal.pone.0079008.

4. Ross BD, Moffat BA, Lawrence TS, et al. Evaluation of cancer therapy using diffusion magnetic resonance imaging. Mol. Cancer Ther. 2003;2:581–587.

5. Moffat BA, Chenevert TL, Lawrence TS, et al. Functional diffusion map: a noninvasive MRI biomarker for early stratification of clinical brain tumor response. Proc Natl Acad Sci USA 2005;102:5524–5529. doi: 10.1073/pnas.0501532102.

6. Morse DL, Galons J-P, Payne CM, Jennings DL, Day S, Xia G, Gillies RJ. MRI-measured water mobility increases in response to chemotherapy via multiple cell-death mechanisms. NMR Biomed 2007;20:602–614. doi: 10.1002/nbm.1127.

7. Badaut J, Ashwal S, Adami A, Tone B, Recker R, Spagnoli D, Ternon B, Obenaus A. Brain water mobility decreases after astrocytic aquaporin-4 inhibition using RNA interference. J Cereb Blood Flow Metab 2011;31:819–831. doi: 10.1038/jcbfm.2010.163.

8. Nilsson M, Englund E, Szczepankiewicz F, van Westen D, Sundgren PC. Imaging brain tumour microstructure. NeuroImage 2018. doi: 10.1016/j.neuroimage.2018.04.075.

9. Lampinen B, Szczepankiewicz F, Mårtensson J, van Westen D, Sundgren PC, Nilsson M. Neurite density imaging versus imaging of microscopic anisotropy in diffusion MRI: A model comparison using spherical tensor encoding. NeuroImage 2017;147:517–531. doi: 10.1016/j.neuroimage.2016.11.053.

10. Novikov DS, Kiselev VG, Jespersen SN. On modeling. Magnetic Resonance Medicine 2018;79:3172–3193. doi: 10.1002/mrm.27101.

11. Jensen JH, Helpern JA, Ramani A, Lu H, Kaczynski K. Diffusional kurtosis imaging: the quantification of non-gaussian water diffusion by means of magnetic resonance imaging. Magn Reson Med 2005;53:1432–1440. doi: 10.1002/mrm.20508.

12. Delgado AF, Fahlström M, Nilsson M. Diffusion kurtosis imaging of gliomas grades II and III-a study of perilesional tumor infiltration, tumor grades and subtypes at clinical presentation. Radiology and … 2017. doi: 10.1515/raon-2017-0010.

13. Raab P, Hattingen E, Franz K, Zanella FE, Lanfermann H. Cerebral Gliomas: Diffusional Kurtosis Imaging Analysis of Microstructural Differences. 2010:1–2.

14. Van Cauter S, Veraart J, Sijbers J, et al. Gliomas: diffusion kurtosis MR imaging in grading. 2012;263:492–501. doi: 10.1148/radiol.12110927.

15. Szczepankiewicz F, van Westen D, Englund E, Westin C-F, Ståhlberg F, Lätt J, Sundgren PC, Nilsson M. The link between diffusion MRI and tumor heterogeneity: Mapping cell eccentricity and density by diffusional variance decomposition (DIVIDE). NeuroImage 2016;142:522–532. doi: 10.1016/j.neuroimage.2016.07.038.

16. Stejskal EO, Tanner JE. Spin diffusion measurements: Spin echoes in the presence of a time-dependent field gradient. J Chem Phys 1965;42:288–292.

17. Wong EC, Cox RW, Song AW. Optimized isotropic diffusion weighting. Magn Reson Med 1995;34:139–143.

18. Cory DG, Garroway AN, Miller JB. Applications of spin transport as a probe of local geometry. Polym. Preprints 1990;31:149.

19. Westin CF, Szczepankiewicz F, Pasternak O, Özarslan E, Topgaard D, Knutsson H, Nilsson M. Measurement Tensors in Diffusion MRI: Generalizing the Concept of Diffusion Encoding. Med Image Comput Comput Assist Interv 2014;8675:209–216.

20. Westin C-F, Knutsson H, Pasternak O, et al. Q-space trajectory imaging for multidimensional diffusion MRI of the human brain. NeuroImage 2016;135:345–362. doi: 10.1016/j.neuroimage.2016.02.039.

21. Eriksson S, Lasič S, Nilsson M, Westin C-F, Topgaard D. NMR diffusion-encoding with axial symmetry and variable anisotropy: Distinguishing between prolate and oblate microscopic diffusion tensors with unknown orientation distribution. J Chem Phys 2015;142:104201. doi: 10.1063/1.4913502.

22. Mitra PP. Multiple wave-vector extensions of the NMR pulsed-field-gradient spin-echo diffusion measurement. Phys Rev B 1995;51:15074–15078.

23. Eriksson S, Lasič S, Topgaard D. Isotropic diffusion weighting in PGSE NMR by magic-angle spinning of the q-vector. J Magn Reson 2013;226:13–18. doi: 10.1016/j.jmr.2012.10.015.

24. Lasič S, Szczepankiewicz F, Eriksson S, Nilsson M, Topgaard D. Microanisotropy imaging: quantification of microscopic diffusion anisotropy and orientational order parameter by diffusion MRI with magic-angle spinning of the q-vector. Frontiers in Physics 2014:1–35.

25. Shemesh N, Jespersen SN, Alexander DC, et al. Conventions and nomenclature for double diffusion encoding NMR and MRI. Magn Reson Med 2016;75:82–87. doi: 10.1002/mrm.25901.

26. Jespersen SN, Lundell H, Sønderby CK, Dyrby TB. Orientationally invariant metrics of apparent compartment eccentricity from double pulsed field gradient diffusion experiments. NMR Biomed 2013. doi: 10.1002/nbm.2999.

27. Szczepankiewicz F, Lasič S, van Westen D, Sundgren PC, Englund E, Westin C-F, Ståhlberg F, Lätt J, Topgaard D, Nilsson M. Quantification of microscopic diffusion anisotropy disentangles effects of orientation dispersion from microstructure: applications in

healthy volunteers and in brain tumors. NeuroImage 2015;104:241–252. doi: 10.1016/j.neuroimage.2014.09.057.

28. Sjölund J, Szczepankiewicz F, Nilsson M, Topgaard D, Westin C-F, Knutsson H. Constrained optimization of gradient waveforms for generalized diffusion encoding. J Magn Reson 2015;261:157–168. doi: 10.1016/j.jmr.2015.10.012.

29. Szczepankiewicz F. Whole-brain diffusional variance decomposition (DIVIDE): Demonstration of technical feasibility at clinical MRI systems. 2017:1–13.

30. de Almeida Martins JP, Topgaard D. Two-dimensional correlation of isotropic and directional diffusion using NMR. Physical Review Letters 2016. doi: 10.1103/PhysRevLett.116.087601.

31. Lätt J, Nilsson M, Brockstedt S, Wirestam R, Ståhlberg F. Bias free estimates of the diffusional kurtosis in two minutes: Avoid solving the kurtosis tensor. Proc Intl Soc Mag Reson Med 2010:3972.

32. Hansen B, Lund TE, Sangill R, Jespersen SN. Experimentally and computationally fast method for estimation of a mean kurtosis. Magn Reson Med 2013;69:1754–1760. doi: 10.1002/mrm.24743.

33. Kiselev VG. Fundamentals of diffusion MRI physics. NMR Biomed 2017;30. doi: 10.1002/nbm.3602.

34. Nilsson M, Szczepankiewicz F, Lampinen B, Ahlgren A, de Almeida Martins JP, Lasič S, Westin C-F, Topgaard D. An open-source framework for analysis of multidimensional diffusion MRI data implemented in MATLAB. Proc Intl Soc Mag Reson Med 2018:1–3.

35. Tournier JD, Calamante F, Connelly A. Determination of the appropriate b value and number of gradient directions for high-angular-resolution diffusion-weighted imaging. NMR Biomed 2013;26:1775–1786. doi: 10.1002/nbm.3017.

36. Akkerman EM. Efficient measurement and calculation of MR diffusion anisotropy images using the Platonic variance method. Magn Reson Med 2003;49:599–604. doi: 10.1002/mrm.10365.

37. Hansen B, Lund TE, Sangill R, Stubbe E, Finsterbusch J, Jespersen SN. Experimental considerations for fast kurtosis imaging. Magnetic Resonance Medicine 2016;76:1455–1468. doi: 10.1002/mrm.26055.

38. Ianuş A, Jespersen SN, Serradas Duarte T, Alexander DC, Drobnjak I, Shemesh N. Accurate estimation of microscopic diffusion anisotropy and its time dependence in the mouse brain. NeuroImage 2018;183:934–949. doi: 10.1016/j.neuroimage.2018.08.034.

39. Jones DK, Horsfield MA, Simmons A. Optimal strategies for measuring diffusion in anisotropic systems by magnetic resonance imaging. Magn Reson Med 1999;42:515–525.

40. Klein S, Staring M, Murphy K, Viergever MA, Pluim JPW. elastix: a toolbox for intensity-based medical image registration. IEEE Trans Med Imaging 2010;29:196–205. doi: 10.1109/TMI.2009.2035616.

41. Nilsson M, Szczepankiewicz F, van Westen D, Hansson O. Extrapolation-Based References Improve Motion and Eddy-Current Correction of High B-Value DWI Data: Application in Parkinson's Disease Dementia Cercignani M, editor. PLoS One 2015;10:e0141825. doi: 10.1371/journal.pone.0141825.s002.

42. Chuhutin A, Hansen B, Jespersen SN. Precision and accuracy of diffusion kurtosis estimation and the influence of b-value selection. NMR Biomed 2017;30. doi: 10.1002/nbm.3777.

43. Phillips JJ, Misra A, Feuerstein BG, Kunwar S, Tihan T. Pituicytoma: characterization of a unique neoplasm by histology, immunohistochemistry, ultrastructure, and array-based comparative genomic hybridization. Arch. Pathol. Lab. Med. 2010;134:1063–1069. doi: 10.1043/2009-0167-CR.1.

44. Setsompop K, Cohen-Adad J, Gagoski BA, Raij T, Yendiki A, Keil B, Wedeen VJ, Wald LL. Improving diffusion MRI using simultaneous multi-slice echo planar imaging. NeuroImage 2012;63:569–580. doi: 10.1016/j.neuroimage.2012.06.033.

45. Alexander DC. Axon radius measurement in vivo from diffusion MRI: A feasibility study. 2007:1–8.

46. Lampinen B, Szczepankiewicz F, van Westen D, Englund E, C Sundgren P, Lätt J, Ståhlberg F, Nilsson M. Optimal experimental design for filter exchange imaging: Apparent exchange rate measurements in the healthy brain and in intracranial tumors. Magnetic Resonance Medicine 2017;77:1104–1114. doi: 10.1002/mrm.26195.

47. Szczepankiewicz F, Nilsson M. Maxwell-compensated waveform design for asymmetric diffusion encoding. In:; 2018.

48. Lundell H, Nilsson M, Westin C-F, Topgaard D, Lasič S. Spectral anisotropy in multidimensional diffusion encoding. ISMRM Proceedings 2018.

49. Jespersen S, Olesen JL, Ianuş A, Shemesh N. Implications of nongaussian diffusion on the interpretation of multidimensional diffusion measurements. ISMRM Proceedings 2018.

50. Clark CA, Hedehus M, Moseley ME. Diffusion time dependence of the apparent diffusion tensor in healthy human brain and white matter disease. Magn Reson Med 2001;45:1126–1129.

51. Nilsson M, Lätt J, Nordh E, Wirestam R, Ståhlberg F, Brockstedt S. On the effects of a varied diffusion time in vivo: is the diffusion in white matter restricted? Magn Reson Imaging 2009;27:176–187. doi: 10.1016/j.mri.2008.06.003.

52. Falk Delgado A, Nilsson M, van Westen D, Falk Delgado A. Glioma Grade Discrimination with MR Diffusion Kurtosis Imaging: A Meta-Analysis of Diagnostic Accuracy. 2018;287:119–127. doi: 10.1148/radiol.2017171315.

53. Yao A, Pain M, Balchandani P, Shrivastava RK. Can MRI predict meningioma consistency?: a correlation with tumor pathology and systematic review. Neurosurg Rev 2016. doi: 10.1007/s10143-016-0801-0.

54. Brabec J, Szczepankiewicz F, Englund E, Bengzon J, Knutsson L, Westin C-F, Sundgren

PC, Nilsson M. B-tensor encoding in meningiomas: Comparisons with histology, microimaging and tumor consistency. ISMRM Proceedings 2019.